 \documentstyle[12pt]{article}
\newcommand{\be}{\begin{equation}}
\newcommand{\bea}{\begin{eqnarray}}
\newcommand{\eea}{\end{eqnarray}}
\newcommand{\ee}{\end{equation}}
\newcommand{\lb}{\label}
\begin{document}

\begin{titlepage}
\begin{flushright}
Freiburg THEP-94/30\\
gr-qc/9410029
\end{flushright}
\vskip 1cm
\begin{center}
{\large\bf  SYMMETRIES, SUPERSELECTION RULES, AND DECOHERENCE}
\vskip 2cm
{\bf D. Giulini\footnote{Present address: Center for Gravitational Physics
and Geometry, The Pennsylvania State University, 104 Davey Laboratory,
University Park, PA 16802-6300.}
and C. Kiefer}
\vskip 0.4cm
 Fakult\"at f\"ur Physik, Universit\"at Freiburg,\\
  Hermann-Herder-Str. 3, D-79104 Freiburg, Germany.
\vskip 0.7cm
{\bf H. D. Zeh}
\vskip 0.4cm
 Institut f\"ur Theoretische Physik, Universit\"at Heidelberg,\\
 Philosophenweg 19, D-69120 Heidelberg, Germany.
\end{center}
\vskip 2cm
\begin{center}
{\bf Abstract}
\end{center}
\begin{quote}
We discuss the applicability of the programme of decoherence --
emergence of approximate classical behaviour through interaction
with the environment -- to cases where it was suggested
that the presence of symmetries would lead to exact superselection
rules. For this dis\-cussion it is useful to
make a distinction between
pure symmetries and redundancies, which results from an investigation
into the constraint equations of the corresponding theories.
We discuss, in particular, superpositions of states with different charges,
as well as with different masses, and suggest how the corresponding
interference terms, although they exist in principle,
 become inaccessible through decoherence.
 \end{quote}
\vskip 2cm
\begin{center}
{\em Submitted to Physics Letters A}
\end{center}

\end{titlepage}

It is evident that most superpositions of quantum states do not seem
to be realised in Nature, although they would be allowed by the
 {\em linearity}
of quantum mechanics. Familiar examples are the apparent absence of
superpositions for macroscopic bodies being at different places, or
of superpositions of states with different charges, among many others.
This fact led, in the past, to the introduction of ``superselection
rules" which seem to {\em forbid} the occurrence of such superpositions
in an ad hoc manner \cite{WWW}.
 A postulate, of course, does not yet provide an
explanation, and the question arose about the possible derivability
of such superselection rules. {}From an inspection of algebraic field
theory texts one can get the impression that some of these rules can,
under certain assumptions, be derived rigorously, see e.g.\cite{SW,Ha}.
On the other hand, it has been shown in a wide range of examples
that the influence of the natural environment to a given system
generally leads to the delocalisation of phase relations and thus gives
rise to the classical appearance of the system. This
mechanism of {\em decoherence} occurs naturally and
is an ubiquitous phenomenon \cite{Ze93,Zu91}.

There remain, however, still a few examples where superselection rules
seem to hold exactly. They are connected with the presence of
{\em symmetries} -- well known examples are the superselection rules
for charge, mass (in the nonrelativistic case), or for particles with
different spin.
The question now arises whether it is possible also to explain
{\em these} superselection rules as an effect of the correlation with
the environment. An affirmative answer would demonstrate that the
universality of the superposition principle could be maintained.
In our Letter we shall put forward the point of view that this
can indeed be done. In the case of charge, for example,
classical properties emerge through
the interaction of a charged particle with electromagnetic fields.

One of the central aspects of this point of view, namely the fact that
superselection rules can be physically interpreted
not in an absolute sense, but
only with respect to a ``reference frame"
(the coupling to an appropriate ``measurement device"), was understood
long ago \cite{AS,Mi,Lu}.
There it was pointed out that existing proofs for superselection rules
could not only be applied to operators generating `internal' symmetries
(such as charge), but also to those generating `external' symmetries (such
as [angular] momentum). They should therefore be treated symmetrically,
 leading
to the apparently wrong result that superselection rules should
exist for all locally conserved quantities. But for generators of external
symmetries, like momentum, the flaw in this proof is not difficult to detect.
While formally  correct, it only refers to the momentum of the {\em whole}
 system,
including the measurement apparatus, with respect to an absolute frame.
The relevant quantity is, however, the
momentum {\it relative} to the reference frame defined by the apparatus.
This essential misidentification thus renders the (formally correct) proof
physically irrelevant. The idea put forward in the references cited above is
that this criticism essentially applies to all such `proofs', and that
consequently  there cannot exist any exact superselection rules. Two further
important aspects were, however, neglected in these early considerations:
First, due to the validity of constraint equations such as Gauss' law,
quantum operators such as the electric charge operator acquire a
different status in the theory than, say, angular momentum. (This
difference largely disappears, however, if angular momentum is treated within
general relativity, since there the momentum constraints are
analogous to Gauss' law, see below.)
 Second, the strong effect of the environment, for example the
electromagnetic field in the case of charges, was not taken into account.
 In fact,
in \cite{AS} only bare charges were considered, although the authors clearly
remarked that only a detailed study of the available interactions could show
whether one could indeed maintain quantum coherence.
 In the present investigation
we thus focus our attention on these two aspects.

We start by exploring the significance of symmetries in the discussion
of superselection rules. It will be appropriate to make first
some general remarks and then address specific examples such as
the charge superselection rule.

How do we describe symmetries in a physical theory? Let us assume that
we are given a (classical) configuration space $Q$. We call it symmetric
under a group $K$ if $k(q)\in Q$ for all $k\in K$ and $q\in Q$.
Quantum mechanically, it gives rise to the existence of symmetry
eigenstates, for example in the form of superpositions
\be \vert\Psi^M_I\rangle =
    \sum_K c_K\int dk\ D^{MK}_I(k)U(k)\vert q\rangle \lb{0} \ee
of all states $U(k)\vert q\rangle=\vert k(q)\rangle$
with matrix representations (Wigner coefficients) $D(k)$ and unitary
representations $U(k)$ on Hilbert space ${\cal H}$. It is assumed
that the ``formal states" (that is, points in the classical phase
space, $P$, or rays in ${\cal H}$) label physical states in a faithful
way in the sense that they correspond to different physical states.
There are, of course, important kinematical symmetries, such as
Galilei or Lorentz transformations, which act only on the state space
-- not separately on configuration space.

The theory is furthermore called {\em dynamically} symmetric
with respect to the action of $K$, if a solution $q(t)$ of the
classical dynamical equations require that $k(q(t))$ be also a
solution. In the corresponding quantum theory, stationary solutions
may then be constructed by means of symmetry eigenstates, since the
Hamiltonian will commute with the generators of the symmetry.

In many cases of interest -- and it is these cases we actually consider
here -- one is given a larger ``configuration space" $\bar{Q}$
on which the physical configurations are labeled in a redundant
fashion. The true configuration space is then obtained
by identifying points
via the action of some group $G$, which we call the {\it redundancy} group
of the theory. Because of its interpretation, $G$ must also
define a {\em dynamical} invariance on this enlarged
configuration space. Quantum mechanically one may formally
introduce an enlarged Hilbert space by means of wave functionals
defined on $\bar{Q}$, although the superposition principle
(understood to apply to {\em physical} states) does not require
the resulting formal states to represent physically new states
in this case. The formal degrees of freedom are in fact
immediately eliminated again by imposing a ``constraint"
that admits only superpositions which are symmetric under the
action of the redundancy group.

 All gauge theories exhibit this structure, where the group of
gauge transformations acts on $\bar{Q}$. Specific examples are
Yang-Mills theories, where the
configuration space $\bar{Q}$ is the space of gauge potentials, and general
relativity, where it is given by the space of three-metrics (with
the coordinate freedom still present).
In these nontrivial examples it proves extremely difficult to eliminate the
redundancies explicitly and find a parametrisation of the proper state
space, $S$. One therefore has to work with
an enlarged state space, $\bar{S}$.
In the corresponding quantum field theories, a member of
$\bar{S}$ may be thought of as a wave functional (or ray
in Hilbert space) on the space of gauge
potentials, or the three-metrics, while a member of $S$ is a wave functional
(or ray) on the space of gauge-invariant quantities, such as the magnetic field
in QED, or the three-geometry in quantum gravity.

Such redundancies should not be confused with proper physical
symmetries as defined above. Their interpretation as well as their consequences
are quite different.
Given an action of a symmetry group $K$ on $S$, we cannot expect it to
be defined on $\bar{S}$. But what must happen is that the symmetries
$K$ and the redundancies $G$ merge into a larger group, $\bar{G}$,
which contains $G$ as a normal subgroup so that the quotient $\bar{G}/G$ is
isomorphic to $K$. A special case is $\bar{G}=G\times K$, but this is
 an exception. The group $\bar{G}$ also transforms solution curves to
solution curves. To account for its hybrid character we simply call it the
{\em invariance} group. Now, the problem encountered in gauge theories, or
general relativity, is that we are given $\bar{S}$ and $\bar{G}$, but not
directly $K$.
To find $K$ one has to decide precisely what part $G$ of $\bar{G}$
corresponds to unobservable transformations.  The constraint analyses of those
theories only tell us that $G$ {\em must}
 contain the group generated by the constraint equations (such as Gauss'
law), which is generally strictly smaller than $\bar{G}$, the group of all
gauge transformations.
A certain freedom is thus left as to whether one should also regard
 the remaining
transformations as redundancy transformations or rather as symmetries.
This distinction becomes important for the discussion of the dynamical origin
of superselection rules.
These general remarks will now be explicitly discussed
in the light of physically relevant examples.

We shall first discuss the case of QED and
the charge superselection rule. (The
formal extension to nonabelian theories
is straightforward, although the possible presence of confinement renders
part of the discussion irrelevant.)
 Here, $\bar{Q}$ is the space of vector potentials
(whose components shall be denoted by $A_a$) and
charged spinor fields
over a three-dimensional manifold $M$ (the $t=constant$ slice), and
$\bar{G}$ is the group of all gauge transformations.
Consider first the {\em classical} theory.
If $E^a$ (the components of the electric field strength) denotes
the momentum conjugate to $A_a$, the phase space function
\be Q^{\xi}[E^a, A_a]=
    \int_M d^3x (E^a\partial_a\xi +\rho\xi) \lb{1} \ee
(where $\xi({\bf x})$ is an arbitrary function and $\rho$ is
the charge density)
generates an infinitesimal gauge transformation parametrised by $\xi$ on
arbitrary functions on phase space. Integration by parts yields
\be Q^{\xi}[E^a,A_a]=
    \int_{S_{\infty}}d\sigma n_aE^a\xi -\int_Md^3x\ \xi
     (\partial_aE^a-\rho), \lb{2} \ee
where the surface integral is over $S_{\infty}$, the ``sphere at infinity",
and $n_a$ is the outward pointing normal. On the other hand, the Gauss
constraint of electrodynamics reads
\be {\cal G}\equiv \partial_aE^a -\rho =0. \lb{3} \ee
One immediately recognises from (\ref{2}) that the Gauss constraint
generates {\em asymptotically trivial} gauge transformations, i.e.,
gauge transformations for which the surface integral in (\ref{2})
vanishes. They form the members of our redundancy group $G$.
One also recognises, however, that $Q^{\xi}$ generates {\em additional}
gauge transformations, for example those which possess a constant $\xi$
at spatial infinity (which is sufficient for the finiteness
of energy). If these additional, ``global", gauge transformations
 were considered as redundancies,
one would exclude global states with nonvanishing overall electric charge.

This may in fact be a reasonable cosmological conclusion
for quantum mechanically closed (spatially finite or infinite)
universes which do not have an outside reference frame,
but it appears inappropriate for systems that in principle can interact
with an outside world. In this sense $S_{\infty}$ is meant to be a
boundary {\em in} rather than {\em of} the physical universe. It acts
 like a reference system with respect to which the conserved quantities
of the system under study are defined. This means that it may also act as a
reservoir for these quantities, just like a material spatial reference
system acts as a reservoir for (angular) momentum or a charge capacitor
as a reference system for the conjugate phase \cite{AS}. It is clear that
for this to make sense the system under study must necessarily share
dynamical correlations with the reference frame, that is with physical systems
outside $S_{\infty}$. We stress the relevance
of this point for our understanding of so-called asymptotically
isolated configurations in field theory. Here, in the mathematical
description, one indeed takes the limit where the radius of $S_{\infty}$
goes to infinity. But this is clearly nothing more than an idealisation
that serves to conveniently  replace boundary conditions by falloff conditions.
These configurations are usually not meant as models of the universe!

Since we interpret $S_{\infty}$ as representing a surface {\em within}
the physical universe,
 we shall here regard global gauge transformations as physically
meaningful {\em symmetries}.
We note that $Q^{\xi}$ commutes with ${\cal G}$
on the constraint surface, and
possesses the value
\be Q^{\xi}\vert_{{\cal G}=0} =\xi\int_{S_{\infty}}
    d\sigma n_aE^a \equiv \xi\ Q, \lb{4} \ee
where
\be Q=\int_{S_{\infty}}d\sigma n_aE^a =\int_M d^3x\rho \lb{5} \ee
is the total electric charge, which is conserved in time,
since a flow through $S_{\infty}$ is here excluded. It is an
observable in the formal sense, since it commutes with ${\cal G}$ on the
constraint surface.

 In the quantum theory the above relations remain
formally valid as operator equations. The charge $\rho$ is then
given by $-ie\pi_{\psi}\psi$, where $\psi$ is the spinor field,
and $\pi_{\psi}$ its conjugate momentum.
If quantisation is performed in the functional Schr\"odinger picture
\cite{Ja}, the constraint (\ref{3}) is implemented as a restriction
on physically allowed wave functionals, $\Psi[A_a,\psi]$, as
\be \partial_a\frac{1}{i}\frac{\delta\Psi}{\delta A_a}
     =-ie\psi\frac{\delta\Psi}{\delta\psi}. \lb{6} \ee
This equation expresses the simultaneous invariance of the wave functional
with respect to local gauge transformations of the vector potential
and the spinor field. The wave functional ({\em if} in a charge eigenstate)
 acquires, however, a phase
if the electric charge operator (\ref{5}) is applied:
\be e^{i\hat{Q}}\Psi=e^{ie\xi_{\infty}}\Psi. \lb{7} \ee
Superpositions of states with different charges thus aquire a relative phase
which involves the value of $\xi$ on $S_{\infty}$.

It is important to note that the total charge within $S_{\infty}$
{\em commutes} with all (quasi) local observables, i.e. observables which
are restricted to subsystems within and suitably bounded away from
$S_{\infty}$ \cite{SW}\footnote{The main part of \cite{SW} is concerned
with the demonstration that the naive calculation of this result with the
help of the above  equations is also valid in local, covariant
gauges, which is necessary to apply the Haag-Kastler formalism of
algebraic field theory.}. The ``state" $\Psi$ in Eq. (8) has therefore
to be understood cosmologically as the decohered {\em component}
of an assumed ``island universe" (which has no sources outside).
 This implies an alternative formulation of
superselection rules, namely that superpositions of different charge,
albeit not forbidden, are indistinguishable from mixed states
{\em with respect to} the chosen class of observables \cite{La}.
 In the algebraic approach to quantum field
theory such a class of observables is introduced from the outset as
a ``reasonable choice".  Whereas this might be a formally consistent procedure,
it seems too rigid to accomodate for a physical explanation of superselection
rules, which at least in some cases
(but probably all cases) we believe to be of an approximate nature
only. The a priori choice of local observables might not necessarily be so
``reasonable''  once our interpretation of $S_{\infty}$
as a boundary {\em within} the physical universe is recalled.
On the other hand, insisting on the limit where $S_{\infty}$ is strictly
infinitely far away, one is left with no range of operational
applicability.

Superpositions of different charge states can be distinguished from
a corresponding mixture if a non-commuting variable is available at
``spatial infinity". This prevents the charge to lie in the centre of
the algebra of the chosen observables.
As can be seen from (\ref{5}), this would necessarily involve
the vector potential at spatial infinity. An example is the
Mandelstam variable
\be \exp\left(ie\int_{-\infty}^x{\bf A}d{\bf s}\right), \lb{8} \ee
where the integration goes over an arbitrary path coming from
``infinity'' to the end point ${\bf x}$.
It has non-vanishing matrix elements between different charge
sectors. If the invariance with respect to ``global" gauge transformations
 is interpreted as a symmetry,
and not as a redundancy, this variable
must correspond to an observable. It is of course a physical question
whether such operators are practically available and whether such
phases at ``infinity"
 can actually be measured, i.e. whether a reference system at such
large distances can be assumed to exist.
(For comparison, in the Kaluza-Klein approach to electric charge, this would
correspond to the measurability of the intrinsic scale of the fifth dimension.)

One can, however, envisage an effective mechanism that prevents us
from seeing such superpositions of charge states.
Because of Gauss' law, Coulomb fields carry information about the
charge {\em at any distance}, thereby decohering a
superposition of different charges in any {\em bounded} subsystem
of an infinite universe. This ``instantaneous" action of decoherence at an
arbitrary distance by means of the Coulomb field gives it the
appearance of a kinematical effect, while it is in fact based on the
dynamical law of charge conservation. Such a time-independent charge
must then always have been ``measured" by the asymptotic field and thus has
always been ``decohered". We emphasise that it was important above that
charge is measurable by putting the sphere
$S_{\infty}$ at any distance on $t=constant$
 which is sufficiently far away. If
Gauss' law did not hold
 (and thus no Coulomb fields did exist), this would not be possible
since one would then only have radiation fields which propagate
along the future light cone.

We also note that there is an interesting connection between the
degrees of freedom at infinity (corresponding to the directions
of symmetry discussed above) and the infrared structure of the theory,
i.e. the r\^{o}le of ``soft photons"
\cite{Ge}. In fact, the presence of these degrees of freedom
determines the infrared structure and demands, in a scattering
process, the presence of
incident soft photons which in turn leads to the cancellation
of infrared divergences in the transition element between in- and
out-states.

The above considerations show that it is
physically more relevant to consider
{\em local} superpositions of charge states.
Let us therefore consider a ``universe" with charge $Q$,
\be \vert\Psi\rangle =\int dq f(q)\vert q\rangle\otimes
    \vert Q-q\rangle \lb{9} \ee
-- either the closed Universe with $Q=0$, or an island universe
that is decohered by its Coulomb field. Eq. (10)
 is an eigenstate of the charge operator for the total system,
with eigenvalue $Q$, but where
 the respective subsystems
 are not in charge eigenstates. Of course, a global
gauge transformation does no longer lead to different phases for the various
components of such a state, but only to one global phase factor
as in (\ref{7}). Consequently, one does not have to invoke
a non-commuting variable at ``infinity'' like (\ref{8}) to verify this
superposition.
An example of such a ``local superposition" can be found in the
BCS state of superconductivity,
\[ \psi_{BCS}=\prod_k \left(1+v(k)a_k^{\dagger}a_{-k}^{\dagger}
   \right)\vert 0\rangle. \]
 The {\em relative} phases of
two superconductors (contained in $v(k)$)
 can even be measured in Josephson junctions.

States like (\ref{9}) have in fact been considered in \cite{AS} and
\cite{Mi}. The formal analogy of this state with one that refers to
 angular momentum
 instead of charge led these authors to the conclusion that
local superpositions of charges should be observable precisely as local
superpositions of angular momenta. While this is true in principle,
this argument should be
completed by the observation (which is important quantitatively)
that charged particles interact very efficiently with
electromagnetic fields, leading to decoherence of the local
superposition \cite{KZ}. We do indeed seem to have an
effective mechanism that prevents charge superpositions from being
found: For example,
 thermal radiation interacts with charged
particles through Thomson scattering and leads to strong decoherence
within a few seconds \cite{JZ}. Even more efficient seems to be
the interaction of, say, an electron with its own radiation field.
Ford has estimated the diminishing influence of this field
on the height of amplitudes in an interference fringe
and was able to distinguish the separate decoherence
effects resulting from vacuum fluctuations and photon emission,
respectively \cite{Fo}.
Using the Feynman-Vernon influence
functional method, Barone and Caldeira calculated the decoherence factor,
$D(t)$, for the interference between two electronic wave packets
\cite{BC}. They found, using appropriate approximations, that
$D(t)$ approaches a time-independent
constant for times larger than about $10^{-24}sec$
and that it does not lead to any decoherence effect for electrons in a solid
where the interatomic spacing is of the order of angstroms.
We mention that both calculations (\cite{Fo} and \cite{BC})
treat the electron non-relativistically.
In addition to the fine structure constant, $D$ contains a cutoff
in the number of field modes, which is assumed to be given roughly
 by the inverse de~Broglie wavelength of the electron.
  One can, however, easily derive from \cite{BC}
 that decoherence must be very efficient if the electronic
wave packets are separated more than about ten \AA.
This seems to be in conflict with experiment, since
effects of coherent wave packets have been observed over a distance
of about 4600 \AA \cite{Sch}.
 This is, of course, still many orders of magnitude below the values
which can be attained for neutral particles such as neutrons.
One must, however, keep in mind that the cutoff
in the number of modes (which was merely guessed)
enters the exponent of $D$ quadratically and may well be lower
than the value given in Ref.~\cite{BC}.  Only a full QED calculation, which as
yet is elusive, would settle this question. Such a calculation
would have to make use of the exact master equation for an electron
obtained by tracing out its radiation field \cite{Di}.

There exist many situations in which decoherence must be expected
to be effective. For example, the superposition principle requires
the existence of quantum states containing macroscopically different
electromagnetic fields. Such superpositions are in fact described by a
general wave functional of QED as it was assumed in (\ref{6}).
Any sufficiently narrow wave packet for a charged particle
feeling different Lorentzian forces from these different fields
would then immediately split into separate components depending on
(and being correlated with) them. The particle can thus be said
to ``measure" the state of the field. Since there are always charged
particles around, no interference effects between macroscopically
different fields can ever be observed. A QED calculation has shown,
for example, that the superposition of two different electric fields
rapidly decoheres if the field values are not too small \cite{Ki92}.
The field therefore {\em seems to be} in one of its classical states
with the corresponding probability.

If it were possible to spatially separate a charge from its anti-charge
 in order to recombine them later on, it would only be due to the
irreversible ``measurements" by the field that ``information" about
their temporary separation (that is, about their dipole moment
and higher moments)
 remained present in the Universe.
In particular, the components of a globally neutral superposition
such as
\[ \phi_+(1)\phi_-(2)+ e^{i\alpha}\phi_-(1)\phi_+(2), \]
with spatially separated states $1$ and $2$, would {\em remain}
decohered if they were brought together. (This example emphasises
the responsibility of the arrow of time, which is present in the
 retarded nature of decoherence for the historical
appearance of our world.)

Thus, although  charge is {\em always} decohered in
 the subsystems of a state like
(\ref{9})
 (as for any conserved additive
quantity \cite{Lu}), the important question is whether these
subsystems
  can be coherently brought together
again after having been spatially separated.
 Decoherence due to Coulomb fields is reversible, while
the influence of radiation fields (``escaping photons") is irreversible
and leads to an upper bound for the coherent separation of charge
pairs.

If one succeeded in reversibly
 preparing the above $\alpha$-dependent superposition,
one could also ``measure" the charge of the state $\phi(1)$
{\em microscopically}, e.g., by
means of an interaction with a spinor field
$\chi$:
\[ \phi_{\pm}(1)\chi_0 \to \phi_{\pm}(1)\chi_{\pm},\]
where $\chi_{\pm}$ indicates charge {\em dependence} of the final state
-- not charge itself.
After recombination of the charges one would be left with an observable
superposition
\[ \phi_{neutral}(\chi_+ +e^{i\alpha}\chi_-) \]
and would thus have verified the existence of the above superposition.

We also note that, interestingly, decoherence is able to distinguish
between action-at-a-distance theories (like the absorber theory by
Wheeler and Feynman) and local field theories: In the former case
decoherence would only be achieved after absorption has taken place,
i.e., usually much later than in the latter case.

Our second example for the connection of
symmetries and superselection rules is concerned with the total mass
of a system.
In the case of many nonrelativistic particles with a Galilei invariant
interaction potential, a superselection rule for the total mass follows
immediately from the fact that the relevant action  of the Galilei
group on state space is given by some nontrivial ray representations
which are inequivalent for different total mass parameters.
For a detailed exposition see e.g. Ref.~\cite{FESt}.
In physical terms, this corresponds to the unobservability of an
overall phase of the wave function, which is the variable conjugate
to the ``mass operator". In our language, the mass operator generates a
redundancy transformation. The situation is different in general
relativity where mass is equivalent to energy, and the conjugate
variable may be interpreted as a clock variable
(i.e., as ``time"). The question about
possible superpositions of, say, different masses of Schwarzschild black
holes is thus equivalent to the question of the observability of clock
variables at spatial ``infinity''.

In order to describe the situation properly we consider, in analogy to the
field theoretic case above, the formulation of canonical general
relativity for asymptotically isolated systems. Recall that in general
relativity the canonical variables are $(h_{ab},\pi^{ab})$,
where $h_{ab}$ is the Riemannian metric on a three-dimensional
manifold $M$ which in the spacetime picture corresponds to a spacelike
slice. The conjugate momentum, $\pi^{ab}$, is in the spacetime
picture essentially given by the extrinsic curvature. The asymptotic
conditions characterising isolated systems imply the existence of
an {\em asymptotic coordinate system} where the metric approaches
 $\delta_{ab}$ plus terms of the order $1/r$.
The phase space functional
\be D^{\xi}[h_{ab},\pi^{ab}]\equiv \int_Md^3x\left(2 \pi^{ab}
    \nabla_a\xi_b + T^{nb}\xi_b\right) \lb{11} \ee
(where $T^{nb}$ denotes the corresponding components of the
energy momentum tensor with $n$ denoting the normal component)
generates infinitesimal diffeomorphisms. As in the case of charge
we perform an integration by parts,
\be D^{\xi}=2\int_{S_{\infty}}d\sigma \pi^{ab}n_a\xi_b
    -\int_M\left(2\nabla_a\pi^{ab}-T^{nb}\right)\xi_b d^3x. \lb{12} \ee
On the other hand, the diffeomorphism constraints are given by
(note the analogy to Gauss' law (\ref{3}))
\be {\cal C}\equiv 2\nabla_a\pi^{ab}-T^{nb}=0, \lb{13} \ee
and are seen from (\ref{12}) to generate the connected component
of asymptotically trivial diffeomorphisms (for which the surface
integral in (\ref{12}) vanishes). This will represent
 our redundancy group $G$.
 In quantum gravity the constraint
(\ref{13}) is implemented as a constraint on wave functionals and
implies that this functional does not change with respect to this
kind of diffeomorphisms. It can, however, acquire a phase by
an asymptotically nontrivial diffeomorphism, or a diffeomorphism
that is not connected with the identity (giving rise to $\theta$-
states, see e.g. \cite{Is},
 \cite{Ki93}).
The configuration space
$\bar{Q}$ is now given
 by the space of these metrics on $M$, while $\bar{G}$ is the whole group
of diffeomorphisms of $M$ which preserve the asymptotic structure for
the metric. The symmetry group $K=\bar{G}/G$ thus still contains
those diffeomorphisms that preserve the asymptotic form for the metric,
that is, the euclidean group of translations and rotations.
In the case of rotations, for example, one finds
\be D^{\xi}\vert_{{\cal C}=0} =\omega^aJ_a, \lb{14} \ee
where $\omega^a$ and $J_a$ are the components of angular velocity
and angular momentum, respectively. This is the analogous expression
to (\ref{4}). A similar expression follows for the total momentum.
Thus, in general relativity, total momentum and angular momentum
are expressible as asymptotic charges which are determined by surface
integrals over arbitrarily large spheres. Again, these charges
commute with the diffeomorphism constraint and thus correspond to
formal observables.

The question whether these charges are observables is connected with
the question about the physical realisability of the asymptotic
coordinate system. It is plausible that a rotation
with respect to some ``absolute space"
is meaningless and thus defined only with respect to
some {\em physically} defined asymptotic space.
(This, of course, is reminiscent of Mach's criticism
of Newtonian space concepts.) Superpositions of different angular
momenta are then possible and no superselection rule holds
for them. Note that
there are well known solutions with non-vanishing asymptotic charges,
most notably the Kerr black hole. If no physically defined
asymptotic space were available, only states with
vanishing angular momentum would be allowed. This feature is
already present in purely relational and reparametrisation invariant
theories of nonrelativistic mechanics \cite{BB}.

A discussion of a superselection rule for mass in general relativity
has to address the r\^{o}le of energy.
 General relativity contains, in addition to the
diffeomorphism constraint (\ref{11}), a Hamiltonian constraint
which classically generates reparametrisations in time. In the
asymptotically flat case it contains the total energy of the system
as a surface integral at ``infinity'',
\be E= \frac{1}{16\pi G}\int_{S_{\infty}}
   d\sigma n_a(h_{ab,b}-h_{bb,a}), \lb{15} \ee
and the momentum conjugate to a clock variable at ``infinity''
\cite{RT}. One may then, as in Eq.~(\ref{0}), construct a
superposition of all possible pointer positions of this clock,
arriving in this way at an energy eigenstate.
The unobservability of this ``global time" would immediately lead
to a superselection rule for the energy \cite{PW}, in the
same way as the unobservability of a global phase would lead to
a superselection rule for the total charge. In analogy to
(\ref{9}), time emerges intrinsically in the form of correlations
between subsystems such as clocks \cite{PW,ZeB}.
 If the Universe is closed, the clock variable at infinity is absent, since
no ``outside region" is available
as a reference system.

We also note an interesting analogy between the Coulomb potential
and the lapse function. In the four dimensional picture
with Newtonian approximation, the square of the lapse function
becomes $1+2U$, where $U$ is the Newtonian potential. Like the
Coulomb potential, it is determined after gauge fixing by an elliptic
equation and thus propagates ``instantaneously" to infinity.
This explains why the energy in (\ref{15}) is given by a surface
integral, in analogy to the charge (\ref{5}).

 The analogy between charge and angular
momentum etc. is, however, not complete. In the case of charge
it is the global eigenvalue that decoheres. In the case of translations,
 the variable conjugate
to momentum, namely position, becomes ``classical" in this sense,
while the nonabelian character of rotations leads to the classical
 orientation of the rotation axis (see \cite{ZeB}).

For local superpositions of macroscopic masses, decoherence should
occur through interaction with the gravitational field as well as
with matter fields. As far as the interaction with retarded fields
is concerned, one has to take into account the small
interaction strength. (One would expect a factor
$Gm^2/5c^5$ instead of $2e^2/3c^3$ to occur in the decoherence
factor of \cite{BC}).
Matter, however, ``measures" the geometry very efficiently \cite{Ki92}.

We finally emphasise that we expect other examples to fit into
this scheme as well.
 One example is provided by ``large" gauge transformations
and their associated $\theta$- states
\cite{Ki93}. In QCD, superpositions would
lead to superpositions of neutron states with a different electric
dipole moment. One would again expect that the interaction with
electromagnetic fields would decohere this superposition.

\vskip3mm

{\em Note added}: After completion of this work the preprint
 ``Conservation Laws in the Quantum Mechanics of
Closed Systems", by J. B. Hartle, R. Laflamme, and D. Marolf,
has appeared on the bulletin board
(gr-qc/9410006). It discusses the connection of
superselection rules and symmetries in the framework of consistent
histories. Their result, that decoherence is exact as a
consequence of the coupling to long-range fields,
corresponds to our case where the spheres in the integrations
of (3) and (12) lie {\em strictly} at infinity, thus leading to an exact
superselection rule.

\vskip5mm

\begin{center}
{\bf Acknowledgements}
\end{center}
We are grateful to Heinz-Dieter Conradi, Erich Joos,
Joachim Kupsch and Ion-Olimpiu Stamatescu for many stimulating
discussions and critical comments. We also thank the FESt for support.



\begin{thebibliography}{99}

\bibitem{WWW} G. C. Wick, A. S. Wightman, and E. P. Wigner,
              Phys. Rev. {\bf 88}, 101 (1952).

\bibitem{SW} F. Strocchi and A. S. Wightman, Journ. Math. Phys. {\bf 15},
             2198 (1974).

\bibitem{Ha} R. Haag, {\em Local quantum physics}, Springer, Berlin,
             1992;\\ S. S. Horuzhy, {\em Introduction to
             Algebraic Quantum Field Theory}, Kluwer, Dordrecht, 1990.

\bibitem{Ze93}
H. D. Zeh, Found. Phys. {\bf 1}, 69 (1970);
           Phys. Lett. A {\bf 172}, 189 (1993).

\bibitem{Zu91} W. H. Zurek, Phys. Rev. D {\bf 24}, 1516 (1981).

\bibitem{AS} Y. Aharonov and L. Susskind, Phys. Rev.
 {\bf 155}, 1428 (1967).

\bibitem{Mi} R. Mirman, Phys. Rev. D {\bf 1}, 3349 (1970).

\bibitem{Lu} E. Lubkin, Ann. Phys. (N.Y.) {\bf 56}, 69 (1970).

\bibitem{Ja} R. Jackiw, in {\em Field Theory and Particle
  Physics}, edited by O. Eboli, M. Gomes, and A. Santano
  (World Scientific, Singapore, 1988); C. Kiefer and A. Wipf,
  {\em Functional Schr\"odinger equation for fermions in
   external gauge fields}, to appear in Annals of Physics.

\bibitem{La} N. P. Landsman, Int. Journ. Mod. Phys. A {\bf 6}, 5349 (1991).

\bibitem{Ge} J.-L. Gervais and D. Zwanziger, Phys. Lett. B {\bf 94},
             389 (1980).

\bibitem{KZ} O. K\"ubler and H. D. Zeh,
             Ann. Phys. (N.Y.) {\bf 76}, 405 (1973).

\bibitem{JZ} E. Joos and H. D. Zeh, Z. Phys. B {\bf 59}, 223 (1985).

\bibitem{Fo} L. H. Ford, Phys. Rev. D {\bf 47}, 5571 (1993).

\bibitem{BC} P. Barone and A. Caldeira, Phys. Rev. A {\bf 43},
             57 (1991).

\bibitem{Sch} H. Schmid, in {\em Proceedings of the 8th
  European Congress on Electron Microscopy}, edited by
  A. Csan\'{a}dy et al. (Program Committee, Budapest, 1984).

\bibitem{Di} L. Di\'{o}si, Found. Phys. {\bf 20}, 63 (1990).

\bibitem{Ki92} C. Kiefer, Phys. Rev. D {\bf 46}, 1658 (1992).

\bibitem{FESt} D. Giulini, E. Joos, C. Kiefer, J. Kupsch,
               I.-O. Stamatescu, and H. D. Zeh, {\em Decoherence
               and the Appearance of a Classical World in Quantum
               Theory}, Chapter V (in preparation).

\bibitem{Is} C. J. Isham, Phys. Lett. B {\bf 106}, 188 (1981).

\bibitem{Ki93} C. Kiefer, Phys. Rev. D {\bf 47}, 5414 (1993).

\bibitem{BB} J. B. Barbour and B. Bertotti, Proc. R. Soc. Lond. A {\bf 382},
             295 (1982).

\bibitem{RT} T. Regge and C. Teitelboim, Ann. Phys. (N.Y.) {\bf 88}, 286
(1974).

\bibitem{PW} D. N. Page and W. K. Wootters, Phys. Rev. D {\bf 27},
    2885 (1983).

\bibitem{ZeB} H. D. Zeh, in {\em Stochastic evolution of quantum states
    in open systems and measurement processes}, edited by L. Di\'{o}si
    and B. Lucacz (World Scientific, Singapore, 1994).


\end{thebibliography}
\end{document}